\documentclass[conference]{IEEEtran}
\IEEEoverridecommandlockouts
\usepackage{graphicx}
\usepackage{subfigure}
\usepackage{caption}
\usepackage{algorithm}
\usepackage{algcompatible}
\usepackage{booktabs}
\usepackage{amsmath}
\usepackage{amsthm}
\usepackage{cite}

\def\BibTeX{{\rm B\kern-.05em{\sc i\kern-.025em b}\kern-.08em
    T\kern-.1667em\lower.7ex\hbox{E}\kern-.125emX}}

\begin{document}

\title{A Multi-Threading Algorithm for Constrained\\
Path Optimization Problem on Road Networks}

\author{\IEEEauthorblockN{Kousik Kumar Dutta}
\IEEEauthorblockA{
\textit{Indian Institute of Technology Ropar}\\
\textit{Department of CSE}\\
Rupnagar, India\\
kousik.21csz0004@iitrpr.ac.in}
\and
\IEEEauthorblockN{Ankita Dewan}
\IEEEauthorblockA{
\textit{Indian Institute of Technology Ropar}\\
\textit{Department of CSE}\\
Rupnagar, India\\
2017csz0012@iitrpr.ac.in}
\and
\IEEEauthorblockN{Venkata M. V. Gunturi}
\IEEEauthorblockA{
\textit{Indian Institute of Technology Ropar}\\
\textit{Department of CSE}\\
Rupnagar, India\\
gunturi@iitrpr.ac.in}
}

\maketitle

\begin{abstract}
The constrained path optimization (CPO) problem takes the following input: (a) a road network represented as a directed graph, where each edge is associated with a ``cost'' and a ``score'' value; (b) a source-destination pair and; (c) a budget value, which denotes the maximum permissible cost of the solution. Given the input, the goal is to determine a path from source to destination, which maximizes the ``score'' while constraining the total ``cost'' of the path to be within the given budget value. CPO problem has applications in urban navigation. However, the CPO problem is computationally challenging as it can be reduced to an instance of the arc orienteering problem, which is known to be NP-hard. The current state-of-the-art algorithms for this problem are essentially serial in nature and cannot take full advantage (i.e., achieve good load balance) of the increasingly available multi-core systems to solve a CPO query. Our proposed parallel algorithm (with its intelligent task-assignment scheme) achieves both superior solution quality and very low execution times (via good load balancing). Moreover, our approach is also able to demonstrate an almost linear speed-up with an increase in the number of cores.
\end{abstract} 


\section{Introduction}
\label{intro}
The problem of finding a path in road networks has been of great importance. Given the significance of the problem area, several researchers (e.g., \cite{tdrive, jing1998hierarchical,Demiryurek2011,Gunturi2015,byang18,Delling2008hb}) have explored it from different aspects. Among these, the most fundamental is computing a path between a source and destination under a given preference metric. The preference metric of choice has typically been the minimization of distance (e.g., \cite{jing1998hierarchical}), time (e.g., \cite{Demiryurek2011,byang18}), or fuel (e.g., \cite{yanli2018}).

However, the increasing proliferation of mobility-based Big Data \cite{mckinseynew,alireem2015} enables one to ask for much more nuanced routing queries such as: ``Determine the path which goes through wide roads while constraining the total length of the path to be at most 5km longer than the shortest path?'' or ``Determine a path which can be easily followed (refer \cite{ramneekdexa}) while constraining the total length of the path to be at most 5km longer than the shortest path.'' The central theme of both these queries is that they have the notion of both an optimizing metric and a constraining metric. Here the aim is to determine a path that maximizes on the optimizing metric (e.g., road width, navigability) along with constraining the path according to the constraining metric (e.g., distance). We refer to such problems as \emph{Constrained Path Optimization (CPO)} problems. 

\noindent \textbf{Importance of Problem:} \emph{Constrained Path Optimization} problems have recently gained interest in the database system community (e.g., \cite{ramneekdexa,ramneekj,shahabi2015,shahabi2017}) from the perspective of developing scalable solutions for some ``instantiations'' of the constrained path optimization problem. For instance, in \cite{shahabi2015,shahabi2017}, the goal was to determine a path that maximized the scenery (i.e., it has more beautiful viewpoints) while constraining the total length of the path in terms of distance. Whereas \cite{ramneekdexa,ramneekj} focused on maximizing the ``navigability'' of the route (in terms of easily identifiable landmarks or easily navigable roads) while constraining the total distance traveled. Thus, it is conceivable that several other potentials ``instantiations'' of the constrained path optimization problem can come forward in the modern age of Big Data, where different road parameters (e.g., road type, road quality, etc.) are increasingly being recorded. 

\subsection{Computational Challenges} 
\label{chal}
An instance of the constrained path optimization (CPO) problem can be reduced to an instance of the arc orienteering problem (AOP), which is known to be an NP-hard problem \cite{exact1, approximation1, Chekuri2005}. Moreover, any typical urban road network would have hundreds of thousands of road segments and road intersections. Thus, scalability is vital for any potential algorithm for the CPO problem. More specifically, we believe that for any navigational system based on the CPO problem to be used in real-life, the underlying path-finding algorithm should have a running time of at most a few seconds.

\noindent \textbf{Challenges in adapting minimization based approach for CPO problem:} There are many noted works done in the area of minimization of preference metric (e.g., \cite{cola,Demiryurek11,foresthop}). However, it is important to note that these approaches cannot be easily modified to solve the CPO problem. A straightforward approach to reducing a maximization problem such as the CPO problem into a minimization problem is to change the sign of score value (i.e., make it negative). However, given that most of the minimization-based approaches (e.g., \cite{cola,Demiryurek11,foresthop}) inherently use a variant of Dijkstra's algorithm for optimization, they would not be able to work with negative score values. Another approach is to use reciprocals of the score values on edges, i.e., replace each score value $x$ with $\frac{1}{x}$, and then use the minimization-based approach. However, with this reduction, we cannot get a one-to-one mapping (with the original maximization-based CPO problem instance). The challenge of this reciprocal-based approach was also detailed in \cite{ramneekj}.

\subsection{Our Contributions} 
This paper makes the following contributions:\\
\noindent (a) To the best of our knowledge, we believe this paper is the first to explore the challenges of developing a parallel approach for the Constrained Path Optimization (CPO) problem on road networks. 
\noindent (b) We propose a novel parallel approach, called \emph{Parallel-Spatial-RG} algorithm, for the CPO problem on road networks. \emph{Parallel-Spatial-RG} algorithm intelligently performs task assignment and obtains good CPU utilization while avoiding deadlocks. 

\noindent (c) We experimentally evaluate the \emph{Parallel-Spatial-RG} algorithm using real road networks and compare it with existing alternative techniques for the CPO problem. 

\noindent (d) Our experimental results obtained significantly better solution quality than the current heuristic algorithm \cite{ramneekdexa,ramneekj} while maintaining comparable running times.

\noindent (e) Our experimental results also showed that our proposed algorithm was able to \textbf{load balance and demonstrate an almost linear speed-up} with an increase in the number of cores.\\

\noindent \textbf{Outline:} The rest of the paper is organized as follows: Section \ref{bc} presents some basic concepts and formally defines the CPO problem. Section \ref{pa} presents our proposed approach. We evaluate our proposed approach and compare it with alternatives in Section \ref{exp}. Finally, we conclude this paper in Section \ref{con}.

\section{Related Work}
\label{rel-work}
Research literature most relevant to our problem consist of the following: (a) work done in the area of theoretical computer science (e.g., \cite{Chekuri2005}), (b) work done in the area of database systems (e.g., \cite{ramneekdexa,shahabi2015,shahabi2017,ramneekj}), (c) work done in the area of parallel algorithms for shortest paths (e.g., \cite{dmsr,subgcenteric2,dsChakaravarthy}), (d) work done in the area of constrained shortest path problem (e.g., \cite{cola,cola-gpu,foresthop,old-csp}).

Researchers from theoretical computer science have focused on developing approximation algorithms for the orienteering problem. Note that orienteering problem can be reduced to arc orienteering problem \cite{GAVALAS2015313}. Researchers \cite{Chekuri2005} proposed a quasi-polynomial algorithm that yields a $O(\log OPT)$ approximation for the orienteering problem. While the approximation ratio is impressive, the execution time of this algorithm is very high due to its high time complexity. Our adaptation of \cite{Chekuri2005} for the CPO problem yielded a running time in hours (or in some cases, days) on a typical road network dataset, which is not feasible in the real world.

Database system researchers have been working on variants of the CPO problem (e.g., \cite{ramneekdexa,shahabi2015,shahabi2017,ramneekj}) from a scalability perspective. Their approaches already have running time in milliseconds, and parallelizing them can not produce a better solution quality. In contrast, our experimental analysis shows that our proposed approach obtains a significantly better solution quality while maintaining comparable running times (more details in Section \ref{exp}).

Research work done in the area of parallel algorithms for shortest paths (e.g., \cite{dmsr,subgcenteric2,gunturi2019,dsChakaravarthy}) has focused on developing path-finding algorithms that optimize routes based on a single preference metric (e.g., distance or time). Hence, these cannot be used for our CPO problem, which has an inherently different computational structure as it uses more than one preference metric.

Lastly, it is important to acknowledge the work done in the area of constraint shortest path (CSP) problem \cite{cola,cola-gpu,foresthop,old-csp} while proposing a solution for the CPO problem. The CSP problem aims to determine the shortest path between a source-destination pair subject to certain constraints. Note that CSP problem is a \emph{minimization based problem} and moreover the existing techniques \cite{cola,cola-gpu,foresthop,old-csp} are based on Dijkstra's algorithm. Therefore, as mentioned in Section \ref{chal}, we cannot use them to solve a CPO problem instance.  

\section{Basic Concepts}
\label{bc}
\noindent \textbf{Road network:} A road network is represented as a directed graph $G=(V,E)$. Here, nodes (set $V$) represent the road intersections, whereas directed edges (set $E$) represent the road segments. Each edge in $E$ is associated with a notion of a cost value ($> 0$) and a score value ($\geq 0$). In this paper, the cost of an edge $e=(x,y)$ corresponds to the euclidean distance between $x$ and $y$.  

\noindent \textbf{Optimizing Metric ($\Gamma()$):} Given any directed path $P_i$, $\Gamma(P_i)$ returns the total ``score'' collected by $P_i$. In its simplest form, $\Gamma(P_i)$ can be defined as a sum of the score values of all the edges constituting the path $P_i$. 

\noindent \textbf{Constraining Metric ($\Phi()$):} Given any directed path $P_i$, $\Phi(P_i)$ returns the total ``cost'' consumed by $P_i$. We define $\Phi(P_i)$ as the sum of the cost values of all the edges constituting the path $P_i$. 

\subsection{Problem definition}
\noindent \textbf{Input:} consists of the following: 

\noindent (1) A road network, $G=(V,E)$, where each node $v \in V$ is associated with certain spatial coordinates.  

\noindent (2) A source $s$ $\in$ $V$ and a destination $d$ $\in$ $V$.

\noindent (3) A positive value $overhead$ corresponds to the maximum permissible cost allowed over the cost of the minimum cost path from $s$ to $d$. This paper uses the term $budget$ to denote the sum of overhead and the cost of the minimum cost path between $s$ and $d$. 

\noindent \textbf{Output:} A directed path $P*$ between $s$ and $d$. 

\noindent \textbf{Objective function:} Maximize $\Gamma(P*)$ 

\noindent \textbf{Constraint:}  $\Phi(P*) \leq budget$

\section{Proposed Approach}
\label{pa}
This section details our proposed parallel algorithm for solving the CPO problem. In Section \ref{sec:filter}, we present our Spatial-RG algorithm, inspired by the approximation algorithm proposed in \cite{Chekuri2005} for the orienteering problem. Spatial-RG algorithm has been optimized to solve CPO problem instances on road networks and is much more amenable for a good \textbf{load-balanced} parallelization (details in Section \ref{sec:parallel}). In Section  \ref{sec:parallel}, we develop a parallel version of the Spatial-RG algorithm. 
 
\subsection{Spatial Recursive Greedy Approach for CPO problem}
\label{sec:filter}
The key idea over here is first to find an initial seed path (shortest path) between the given source and destination nodes and then recursively determine its better replacements. The initial seed path can be determined using any standard shortest path algorithm. In our implementation, we used A* algorithm with \textit{euclidean distance} as the heuristic function \cite{gunturibook}. The algorithm continues the recursion up to maximum recursion depth $\theta$ (an input parameter given to the algorithm). \\

\noindent \textbf{Details of the Spatial-RG algorithm:} A pseudocode of the algorithm is presented in Algorithm \ref{alg2}. Each call to the algorithm primarily takes the following input: (i) ``source node'' $u$, (ii) ``destination node'' $v$ and (iii) remaining budget $\beta$. In the first call to the Spatial-RG algorithm, $u$, $v$, and $\beta$ would be set according to the input values given while defining the CPO query to be processed. Thereafter, $u$, $v$, and $\beta$ change during the course of the recursion calls.    

In each recursion, Algorithm \ref{alg2} first iterates (outer loop on line 9) over all edges $e$ which satisfy the following two filters: (a) $\Gamma(e) > 0$ and, (b) $e$ is inside the ellipse formed by $u$ and $v$ as the foci and $\beta$ as the major axis length. It is important to note that the filter proposed in (a) may affect the quality of the final solution. Nevertheless, we use it to gain performance. Moreover, our experiments also reveal that Spatial-RG still has significantly better solution quality than the current heuristics for the CPO problem. In contrast, filter (b) maintains correctness. It was already explained in \cite{ramneekj,shahabi2015} and thus, we do not include its correctness proof. 

Suppose an edge $e=(x,y)\in E$ is considered in the outer loop (line 9). This means that the current path between $u$ and $v$ would be replaced by a path which is the combination of the following three sub-paths: (1) a path $P_1$ between $u$ and $x$, (2) edge $e=(x,y)$ and, (3) a path $P_2$ between $y$ and $v$. Both $P_1$ and $P_2$ are determined recursively inside the inner loop on line 11 as described next. 

The inner loop of Algorithm \ref{alg2} loop over a set of possible budget values. In each iteration of this inner loop, $P_1$ and $P_2$ are determined recursively for different budget values (line 12 and 13). For each pair of $P_1$ and $P_2$ (as returned by their respective recursive calls), we determine $P_{new}$ as $P_1 \cup e \cup P_2$ (joining $P_1$, $e$, and $P_2$ in same order). We store $P_{new}$ if it has a better score value than the original path between $u$ and $v$ (determined on line 1). Thus, at the termination of the inner while loop, we would have the ``best possible'' $P_1$ and $P_2$, which can be attached with $e$ (current edge being considered on the outer loop) to obtain the ``best possible'' replacement for the current path between $u$ and $v$.  

It is important to note that budget values sent into the recursive calls for determining $P_1$ and $P_2$ (line 12 and line 13) are not entirely independent of each other. More precisely, if a budget value of $b$ is given for determine $P_1$, then a maximum value of $\beta-b-\Phi(e)$ budget can be given for determine $P_2$.  

And lastly, the set of feasible budget values ($b$) for the inner while loop range from $b=Euclidean\_Distance(u,x)$ to  $b=\beta - \Phi(e)$ - $Euclidean\_Distance(y,v)$. Correctness of these values can be easily proved by considering the fact \textit{shortest network distance over a graph would always be greater than or equal to the Euclidean distance. Details omitted due to lack of space}.

\begin{algorithm}[ht]
\caption{Spatial-RG Algorithm}
\label{alg2}
\begin{flushleft}
\textbf{Input:} (a) Input graph $G(V,E)$; (b) source node $u$; (c) destination node $v$; (d) Remaining budget $\beta$; (e) current $level$; (f) maximum recursion depth $\theta$.\\
\textbf{Output:} (a) A directed path $P$ between $u$ and $v$ 
\end{flushleft}
\begin{algorithmic}[1]
\STATE $P$ $\leftarrow$ minimum cost path between $u$ and $v$ 
\IF{$\Phi(P) > \beta$}
\STATE Return Null
\ENDIF 
\IF{level = $\theta$}   /*Maximum recursion depth reached*/
\STATE Return $P$
\ENDIF
\STATE $s_p$ $\leftarrow$ $\Gamma(P)$ /*stores value of optimizing metric of $P$ */
\FORALL{edge $e=(x,y)\in E$ with $\Gamma(e) > 0$ and $e$ inside ellipse(u,v,$\beta$)}
\STATE $b\leftarrow$ Euclidean\_Distance(u,x)
\WHILE{$b \leq \beta-\Phi(e)-$Euclidean\_Distance$(y,v)$}
\STATE $P_1 \leftarrow$ Spatial-RG $(u,x,b,level+1)$ 
\STATE $P_2 \leftarrow$ Spatial-RG $(y,v,\beta-b-\Phi(e),level+1)$ 
\STATE $P_{new} \leftarrow P_1 \cup e \cup P_2$ 
\IF{$(P_1$ $\cap$ $P_2) = null$ \& $\Gamma(P_{new}) > s_p$}
\STATE $P \leftarrow P_{new}$ and $s_p$ $\leftarrow$ $\Gamma(P_{new})$
\ENDIF
\STATE $b\leftarrow b+1$
\ENDWHILE
\ENDFOR
\STATE Return $P$
\end{algorithmic} 
\end{algorithm}

\noindent\textbf{Spatial Indexing for implementing the Ellipse pruning:} We use a \emph{Uniform Grid Indexing} \cite{grid} to efficiently determine the edges present in $ellipse(u,v,\beta)$ on line 9 of Algorithm \ref{alg2}. This is done using the concept of spatial range query.

\subsection{Parallel Algorithm for CPO problem}
\label{sec:parallel}
While the Spatial-RG algorithm uses several spatial filtering strategies and spatial indexing for gaining performance, its running time (as seen in our experiments) was still impracticable for meeting the real-world expectation of getting a solution for the CPO problem within a few seconds. To this end, a parallel version of Spatial-RG algorithm was developed, which can harness the increasingly available multi-core systems to \textbf{improve execution time while still maintaining the same solution quality as Spatial-RG.}

This section describes the proposed parallel algorithm \emph{Parallel-Spatial-RG} for the CPO problem. On a close inspection of \emph{Spatial-RG} (Algorithm \ref{alg2}), one may realize that the algorithm has some inherent parallelism at the following three places: (a) Outer loop (For loop on line 9 in Algorithm \ref{alg2}), (b) Recursion calls on line numbers 12 and 13 (in Algorithm \ref{alg2}), (c) Inner While loop (on line 11 in Algorithm \ref{alg2}). In the proposed algorithm, only options (a) and (b) are considered for parallelization. The while loop (option (c)) on line 11 in Algorithm \ref{alg2} is not considered for parallelization as the thread which creates the tasks by unrolling the loop on line 9 in Algorithm \ref{alg2} would have no other work except the creation of tasks corresponding to loop in line 11. As a result, this thread would be sitting idle while other threads undertake the job of iterating over the while loop. We first discuss our proposed technique for parallelizing the outer loop (For loop on line 9 in Algorithm \ref{alg2}), and then the recursion calls on line numbers 12 and 13. We start our discussion with an intuitive naive approach and highlight its limitations. Following that, we describe the proposed \emph{Parallel-Spatial-RG} algorithm.

 \begin{figure}[ht]
    \centering
    \includegraphics[height=30mm,width=90mm]{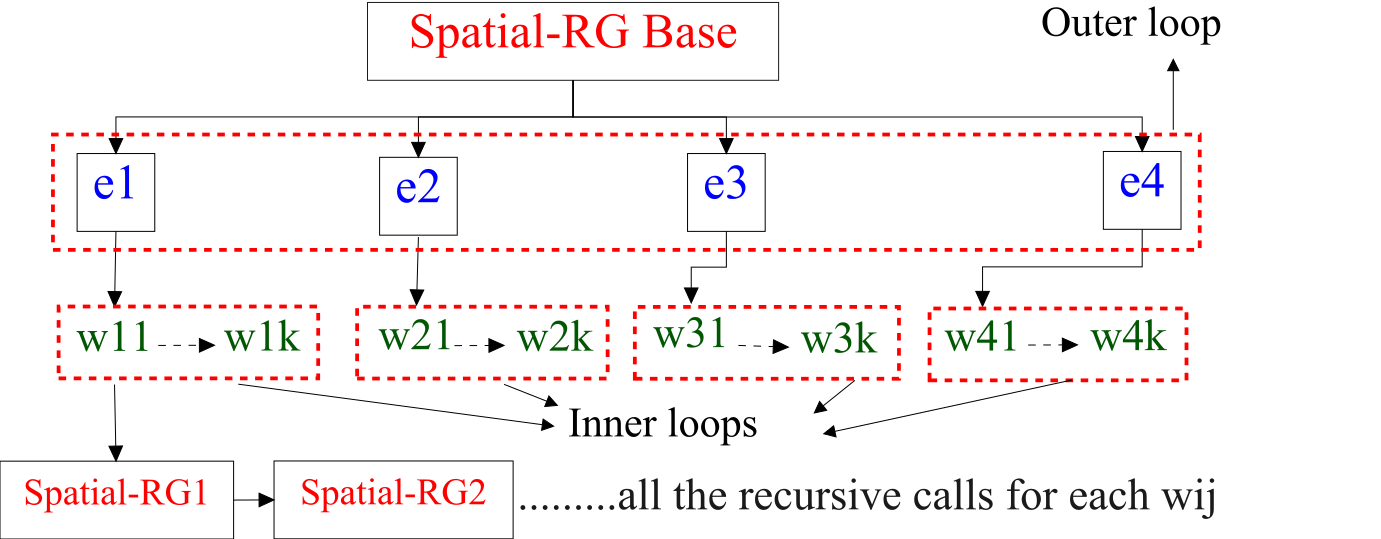}
    \caption{Illustrating the search space of Spatial-RG algorithm.}
    \label{figss}
\end{figure}

\subsubsection{\textbf{Challenges of a Naive Approach (Recursion Unpacking):}}\label{p-naive} Note that \emph{Spatial-RG} (Algorithm \ref{alg2}) has recursion calls inside a while loop. As a result, the search space of the algorithm would be non-linear. More specifically, the search space of the \emph{Spatial-RG} algorithm would be similar to a \emph{tree} structure. And the algorithm would essentially undertake a depth-first traversal of this \emph{tree-structured} search space to obtain a solution.

Figure \ref{figss} illustrates this \emph{tree-structured} search space for a hypothetical instance of Algorithm \ref{alg2}. In this instance, in the first level, the loop on line 9 in Algorithm \ref{alg2} has 4 edges ($e_1, e_2, e_3, e_4$). For each of these edges, the algorithm would first have to execute the while loop (between lines 11 and 20 in Algorithm \ref{alg2}). And inside this while loop, the recursion calls are taking place. 

An intuitive way of parallelizing the \emph{Spatial-RG} algorithm would be to use multiple threads to explore its \emph{tree-structured} search space. As a naive approach, one can employ independent threads to explore the sub-trees (in Figure \ref{figss}) under the root in parallel and then determine the best solution amongst individual solutions obtained in different sub-trees. However, such an approach may not always guarantee good CPU utilization. For instance, consider an 8-core system (capable of running 16 threads). In the search space illustrated in Figure \ref{figss}, one may assign the sub-trees underneath the root to 4 different threads. Such an approach would not be using the remaining 12 threads. Moreover, it may also happen that the work across different sub-trees may not be uniformly distributed due to variation in the properties (density of roads and their scores) of road segments across different spatial locations. Consequently, some threads may finish their work much sooner than others. Therefore, causing even worse CPU utilization.

A logical extension of this naive approach would be to determine an appropriate level in the \emph{tree-structured} search space (at runtime) based on the number of threads available and then continue exploring in parallel. However, this approach also has its limitations. Firstly, on any typical road network, \emph{Spatial-RG}'s search space tree would have a high fan-out factor. In other words, the number of ``nodes'' increases almost exponentially with each level. For instance, level 1 may have 10 nodes, and level 2 may have 80 nodes. Secondly, the search space of \emph{Spatial-RG} (on any CPO problem instance) cannot be pre-determined precisely as it is heavily dependent on the spatial distribution of the edges in the input dataset. 

\subsubsection{\textbf{Details of Parallel-Spatial-RG algorithm:}} As mentioned in the previous subsection, any approach which is solely dependent on unpacking the recursion (followed by simultaneous independent exploration by threads) would have poor CPU utilization. 

To this end, we use the following key strategy: In each recursion call, the current thread is allowed to create further tasks for exploring its designated search space, but those tasks may not always be executed in parallel. More specifically, a separate task is created for each iteration of the outer loop (loop on line 9 in Algorithm \ref{alg2}) and the recursive calls on line numbers 12 and 13 of the \emph{Spatial-RG} algorithm. These tasks are picked up by a thread only if it is idle or free. Otherwise, those tasks are executed serially by the thread which created those tasks (details provided later in the section).

It is important to note that the tasks created by different threads \emph{should not be put in a global job pool}. Such an approach may lead to \emph{deadlock}, as illustrated in the following example. Consider a case where the outer loop (line 9 of Algorithm \ref{alg2}) of \emph{Spatial-RG} and the recursion calls (inside the inner loop) are parallelized. In each recursion call, the current thread would create a separate task for each outer loop iteration. Along with that, the parallelization of the recursion calls also takes place. So, the tasks created previously are further divided into two sub-tasks. All these tasks are put into a global job pool. And whenever a free thread is found, it is assigned a job from the global job pool. 

\begin{figure}[ht]
    \centering
\includegraphics[width=90mm,height=30mm]{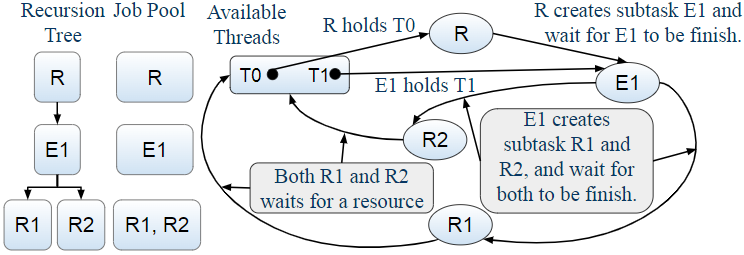}
    \caption{Illustrating deadlock in the naive approach for parallelizing Spatial-RG algorithm.}
    \label{figdl}
    \vspace{-4mm}
\end{figure}

Figure \ref{figdl} illustrates a case of deadlock in such an approach. To simplify the example, we consider a very small instance and a limited number of resources (2 threads). Here in Figure ~\ref{figdl} $R$ is the recursion base. At the start of execution, the $R$ is picked up by thread $T0$. Then $T0$ creates jobs corresponding to all the edges in the outer loop and puts those jobs into the global job pool. In this example, we assume that a single job is created corresponding to a single edge $E1$ (in the outer loop) for the sake of simplicity. Now $T0$ goes into the waiting stage and waits for $E1$ to complete. The job corresponding to $E1$ is then picked by the free thread available in the thread pool $T1$. Following this, $T1$ creates two tasks $R1$ and $R2$ (corresponding to two recursive calls in the inner loop) for $E1$ and puts them into the same global job pool. Now $T1$ also goes into the waiting stage and waits for its recursive tasks ($R1$ and $R2$) to complete. At this stage, there are no free threads in the thread pool. Each thread is waiting for their respective jobs to finish, and jobs in the job pool are waiting for free threads. In Figure ~\ref{figdl}, this situation is explained using two \emph{hold and wait} cycles, which is one of the well-known conditions for \emph{deadlock}. One cycle is formed by the thread pool, $E1$, and $R1$, while the other is formed by the thread pool, $E1$, and $R2$. In each of the cases, $E1$ holds one resource (thread) and waits for its recursive task ($R1$ or $R2$) to complete, and the recursive tasks ($R1$ or $R2$) are waiting for the $R$ or $E1$ to release one of the resources (thread), while $R$ is waiting for $E1$ to be finished. So the algorithm would not proceed further, resulting in a \emph{deadlock}.
  
One may obtain slightly better performance by making each thread hold some tasks for itself and putting the remaining ones in the global job pool. However, this strategy can still not guarantee a deadlock-free execution. Moreover, for this approach to work, one would have to decide the ``optimal number'' of tasks needed to perform serially for each thread before any of their tasks starts execution. Determining this ``optimal number'' precisely would be challenging due to variation in properties (density of roads and their scores) of road segments across different spatial locations.

\begin{algorithm}[ht]
\caption{Parallel-Spatial-RG($u, v, budget, current-level, \theta$):}
\label{alg4}
\begin{algorithmic}[1]
\STATE $P$ $\leftarrow$ minimum cost path between $u$ and $v$.
\IF{$\Phi(P) > budget$}
\STATE Return Null
\ENDIF 
\IF{level = $\theta$}   /*Maximum recursion depth reached*/
\STATE Return $P$
\ENDIF
\STATE Create a job pool $JP$
\STATE Create a result set $RS$ /*$RS$ would store results of jobs in $JP$*/
\FORALL{edge $e=(x,y)\in E$ with $\Gamma(e) > 0$ and $e$ inside the ellipse(u,v,$\beta$)}
\STATE create job for Solve($P, u, v, e, budget, level, \theta$)
\STATE push the job into $JP$.
\ENDFOR
\STATE $\Lambda \leftarrow$ Free threads in the $threadpool$
\FORALL{threads $\lambda_i \in \Lambda$}
\STATE Assign a job from JP to $\lambda_i$ \\
/*On finishing the job, $\lambda_i$ would record its solution in $RS$*/
\ENDFOR
\WHILE{there exists an unassigned job $J_i$ in $JP$}
\IF{there is a free thread $\lambda_j$ in $threadpool$}
\STATE Assign $J_i$ to $\lambda_j$
\ELSE
\STATE the current ``primary'' thread picks the job $J_i$
\ENDIF
\ENDWHILE
\WHILE{there exists an unfinished job in $JP$}
\STATE Wait for completion.
\ENDWHILE
\STATE $P_{new} \leftarrow$ Best solution in $RS$ 
\IF{ $\Gamma(P_{new}) > \Gamma(P)$}
\STATE $P \leftarrow P_{new}$
%\STATE $s_p$ $\leftarrow$ $\Gamma(P*)$
\ENDIF
\STATE Return $P$
\end{algorithmic}
\end{algorithm}

We address the problem of deadlocks through the following strategy: Each thread would maintain its \emph{local job pool}. After creating jobs, the current thread would look for free threads. If no free threads are found, the current thread would start picking up jobs from its job pool (while actively looking for free threads).

\begin{algorithm}[ht]
\caption{Solve($P, u, v, e, \beta, level, \theta$):}
\label{alg5}
\begin{algorithmic}[1]
\STATE $b\leftarrow$ Euclidean\_Distance(u,x) /*edge $e=(x,y)\in E$*/
\WHILE{$b \leq \beta-\Phi(e)-$Euclidean\_Distance$(y,v)$}
\STATE Create a job pool $JP2$
\STATE create a job $k_1$ for Parallel-Spatial-RG$(u,x,b,level+1, \theta)$
\STATE add $k_1$ to $JP2$
\STATE create a job $k_2$ for Parallel-Spatial-RG$(y,v,\beta-b-\Phi(e),level+1, \theta)$
\STATE add $k_2$ to $JP2$
\STATE Create a result set $RS2$ /*$RS2$ would store results of jobs in $JP2$*/
\WHILE{there exists an unassigned job $k$ in $JP2$}
\IF{there is a free thread $\lambda_j$ in $threadpool$}
\STATE Assign $k$ to $\lambda_j$
%\STATE Current thread picks the other job from $JP2$
\ELSE
\STATE Current ``primary'' thread picks the job $k$ from $JP2$
\ENDIF
\ENDWHILE
\WHILE{there exists an unfinished job in $JP2$}
\STATE Wait for completion.
\ENDWHILE
\STATE $P_1 \leftarrow$  path return from job $k_1$.
\STATE $P_2 \leftarrow$  path return from job $k_2$.
\STATE $P_{new} \leftarrow P_1 \cup e \cup P_2$ /*Join $P_1$ and $P_2$ using edge $e$ */
\IF{$(P_1$ $\cap$ $P_2) = null$ \& $\Gamma(P_{new}) > \Gamma(P)$}
\STATE $P \leftarrow P_{new}$
\ENDIF
\STATE $b\leftarrow b+1$
\ENDWHILE
 \STATE Return $P$
\end{algorithmic}
\end{algorithm}

The proposed parallel approach is primarily detailed across Algorithm \ref{alg4} and Algorithm \ref{alg5}. Algorithm \ref{alg5} is used inside Algorithm \ref{alg4}. We initialize Algorithm \ref{alg4} by passing the source node, the destination node, budget, current level (set to $0$), and the maximum depth of recursion allowed. We also create a worker pool (referred to as $threadpool$ in the pseudocodes) to be used in Algorithm \ref{alg4} and Algorithm \ref{alg5} according to the number of cores available. Typically, in modern multi-core processors with hyper-threading technology, we would set the number of workers (threads) to twice the number of cores available.  

Overall, Algorithm \ref{alg4} is structurally similar to the \emph{Spatial-RG} algorithm. Algorithm \ref{alg5} is used inside Algorithm \ref{alg4} to do the work corresponding to the while loop between lines 11--20 in \emph{Spatial-RG}. In each recursion call, Algorithm \ref{alg4} creates a task for each iteration of the outer loop (for loop on line 9 in Algorithm \ref{alg2}). This is done in lines 10--13 of Algorithm \ref{alg4}. Each of these tasks essentially attempts to compute the solution while including a particular edge $e=(x,y)$. With reference to the \emph{Spatial-RG} algorithm (Algorithm \ref{alg2}), this is the work done corresponding to the while loop between lines 11--20. 

The jobs created are put in a job pool $JP$, which is \emph{local} to the current recursion call of Algorithm \ref{alg4}. Jobs from $JP$ are assigned to idle threads. These threads would report their respective results in the result set $RS$, which is shared among them. In our implementation, each thread in $JP$ is assigned a unique location in $RS$. This allows all the threads to access the $RS$ simultaneously and avoids a critical section. Consequently, we get better CPU utilization. If the number of idle threads is less than the number of jobs in $JP$, the remaining jobs are picked up by the currently executing thread (which actually created the job queue) while actively looking for free threads. This is done in lines 10--25 in Algorithm \ref{alg4}. With the intent to simplify the notations, the term ``primary thread'' refers to the thread that creates the job pool in a recursion call. There would be only one ``primary thread'' per call. As mentioned earlier, one can trivially modify this algorithm to hold out one task (from $JP$) for the ``primary thread''. This would give slightly better performance. After all the tasks in $JP$ are completed, the ``primary thread'' chooses the best result from $RS$ and returns it.

As mentioned earlier, Algorithm \ref{alg5} focuses on the while loop between lines $11$--$20$ of the \emph{Spatial-RG} algorithm. It creates tasks for the two recursion calls to Algorithm \ref{alg4} and puts them into a job pool $JP2$. Similar to the previous case, this job pool is created locally by its ``primary thread''. Tasks in $JP2$ are assigned to free threads, which, in turn, would report their solutions in a shared result set. And if no free threads are found, the ``primary thread'' picks up the tasks in $JP2$ for processing.

\noindent \textbf{Generalizing Parallel-Spatial-RG for Minimization:}  Algorithm \ref{alg4} and Algorithm \ref{alg5} can be trivially generalized for a minimization case (e.g., minimize the \#potholes in the path) by making two small changes. First, reverse the ``if'' conditions (between $\Gamma(P_{new})$ and $\Gamma(P)$) on line 30 of Algorithm \ref{alg4} and line 22 of Algorithm \ref{alg5}. Second, remove the restriction of using only the edges $e=(x,y)\in E$ with $\Gamma(e) > 0$ on line 10 of Algorithm \ref{alg4}. In general, we believe that our approach is suitable for both maximization and minimization cases. It is important to note that our generalized version of \emph{Parallel-Spatial-RG} would still not use Dijkstra's styled enumeration (as original \emph{Parallel-Spatial-RG} does not use it) while optimizing the score values. This makes it robust to handle edges with negative scores as well. In fact, dependence on Dijkstra's styled enumeration in the current state of art solutions (e.g., \cite{cola-gpu,cola,foresthop} makes them unsuitable for the CPO problem. As otherwise, one could reverse the sign of score values (i.e., make them negative) and run a minimization-based approach. We plan to investigate this generalization further in our future work and compare it with the current state-of-the-art in CSP problem (e.g., \cite{cola,cola-gpu,foresthop}).

\noindent \textbf{Comment on ForkJoinPool based implementation:} ForkJoinPool internally uses a \emph{work-stealing} based scheduler and gives priority to the jobs created by lower levels in the recursion calls. Such a strategy would avoid deadlocks. However, this scheduler does not allow the programmer to control the threads created within a particular task. As a result, race conditions can not be controlled, which leads to low CPU utilization. Thus, a trivial parallelization of the \emph{Spatial-RG} algorithm using the ForkJoinPool library would not give good CPU utilization due to the creation of race conditions at lines 15, 16, and 17 of Algorithm \ref{alg2}.

\noindent \textbf{Time complexity analysis of Parallel-Spatial-RG: }
In the worst case, an instance of \emph{Parallel-Spatial-RG} algorithm would iterate over all the edges $e=(x,y)\in E$ (in the outer loop), and for each iteration, it would again iterate for $\beta$ times (in the worst case) in the inner loop. Following this, it would have two recursion calls inside the inner loop. Thus, the time complexity for one recursion call is $O(2m\beta)$ ($m$ is the total number of edges present in the network). For a maximum recursion depth of $\theta$, the total time complexity of \emph{Parallel-Spatial-RG} would be $O(2m\beta)^\theta$. Despite the high worst-case time complexity, the combination of spatial filters (Section \ref{sec:filter}) and good CPU utilization (via our parallel approach) helps \emph{Parallel-Spatial-RG} to have low execution time in practice.

\section{Experimental Analysis}
\label{exp}
In this section, we experimentally evaluate \emph{Parallel-Spatial-RG} and compare it with the current state-of-the-art. 

\begin{table}[ht]
	\caption{DATASETS}
	\label{tab::dataset}
	\centering
		\begin{tabular}{|c@{}|c@{}|c@{}|}
		\hline
			\textbf{Road Network } & \textbf{\#Nodes } & \textbf{\#Edges }  \\
			\hline
			Delhi & 52576 & 150488   \\
			Buenos Aires & 263783 & 864408  \\ 
		    London & 285050 & 749382   \\ 
			\hline
		\end{tabular}
\end{table}

\subsubsection{Datasets:} Our experimental analysis is done on three real-world road network datasets from \cite{datacite}. The details of the datasets are briefed in Table \ref{tab::dataset}. In each of these datasets, some edges were selected uniformly at random from across the road network and were assigned (randomly) a score value between 1 and 15. Other edges have a score value of zero. We also varied the number of edges, which had non-zero score values (details provided later).
 
\subsubsection{Candidate Algorithms:} We compare our proposed \emph{Parallel-Spatial-RG} algorithm against the following candidates: \emph{(a) ILS*(CEI)\cite{shahabi2015}, (b) MSWBS\cite{ramneekdexa}}. We also adapted and implemented \cite{Chekuri2005} for our CPO problem. However, it showed an execution time in hours (sometimes even a day). Therefore, we did not include it in our experiments.

\subsubsection{Variable Parameters:} The following parameters were varied in our experiments in subsection \ref{sec:sensitivity}.

\textit{Budget ($\beta$):} Recall that budget has been defined as the sum of overhead and the cost of the shortest path. As the budget increases, the working space of \emph{Parallel-Spatial-RG} also increases.
    
\noindent  \textit{Path Length:} Path length affects the running time of all the candidate algorithms. However, the notion of path length can have multiple interpretations in a weighted graph. To ensure the interpretability of the results, for our experiments, we define the path length as the number of edges in the shortest path between the given source and destination. We reported the average value of $100$ source-destination pairs for each path length range. 
    
\noindent     \textit{Recursion Depth ($\theta$):} Different recursion depths ($\theta=1$, $\theta=2$) were tried to determine an ``optimal'' recursion depth for \emph{Parallel-Spatial-RG}. This ``optimal'' recursion depth makes a good balance between solution quality and running time.

\noindent     \textit{Number of Threads ($n$):} \emph{Parallel-Spatial-RG} is run for different number of threads ($8, 16, 32, 64, 96, 128$) to show the near linear speed-up with an increase in the number of cores. 
    
\noindent     \textit{Number of edges with non-zero score:} The proposed algorithm's runtime depends on the road network's density, more specifically, the density of the edges with non-zero scores values in that road network.

\subsubsection{Metrics measured:} We measured the following in our experiments: \textit{(a) Execution time of the algorithm.} \textit{(b) Score gain over the score of the shortest path.} Score gain is defined as the difference between the total score of the solution obtained by the candidate algorithm and the shortest path. 

\subsubsection{Experimental Setup:} All the algorithms (including the candidate algorithm) were implemented in JAVA11. We used an Ubuntu machine with several Intel Xeon Platinum 8280M cores (total capacity of 128 threads) and 2048GB RAM. Also, note that our processor had a base frequency of 2.70GHz (with a max boost frequency of 4.00GHz).

\begin{figure}[ht]
\centering
\subfigure[comparison of run-time.]{\label{fig:recursion-depth-a}\includegraphics[height=30mm,width=40mm]{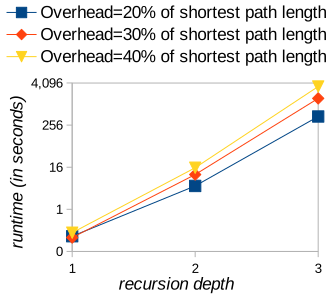}}
\quad
\subfigure[comparison of score gain.]{\label{fig:recursion-depth-b}\includegraphics[height=30mm,width=40mm]{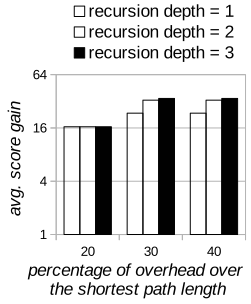}}

\caption{Illustrating performance of Parallel-Spatial-RG for different recursion depths and different overheads. Path lengths $10$--$20$. $40\%$ of edges with non-zero score value. Y-axis is in $log_4$ scale.}
\label{fig:recursion-depth}
\end{figure}

\subsection{Sensitivity Analysis}
\label{sec:sensitivity}
\subsubsection{Varying Recursion Depth ($\theta$):}
\label{sec:recursion-depth}
In this experiment, \emph{Parallel-Spatial-RG} is compared for three different values of recursion depth ($\theta$) ($1$, $2$, \& $3$) at three different values of overhead $30\%$, $40\%$, and $50\%$. The runtime of \emph{Parallel-Spatial-RG} increases exponentially with an increase in the recursion depth. Therefore, we considered a smaller instance of the Delhi road network ($9401$ nodes and $25941$ edges) in which $40\%$ of the edges had a non-zero score value for this experiment. In Fig. \ref{fig:recursion-depth-a} the run-time of \emph{Parallel-Spatial-RG} is shown for different values of $\theta$ at different overheads. Fig. \ref{fig:recursion-depth-b} shows the corresponding score gain. Overall, the figures show that as recursion depth is increased, the execution time of \emph{Parallel-Spatial-RG} increases exponentially. Whereas we only get a very small improvement on the score gain aspect. To this end, we set the recursion depth to $1$ in our remaining experiments.

\subsubsection{Varying number of cores:}
\label{sec:cpu-util}
\begin{figure*}[ht]
\centering
\subfigure[comparison of runtime in Delhi road network.  Y-axis is in $log_4$ scale.]{\label{fig:CPU-utilization-a}\includegraphics[height=45mm,width=57mm]{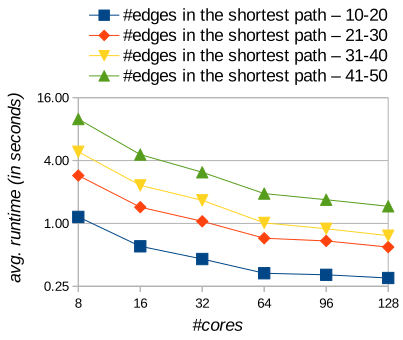}}
\quad
\subfigure[comparison of runtime in Buenos Aires road network.  Y-axis is in $log_4$ scale.]{\label{fig:CPU-utilization-b}\includegraphics[height=45mm,width=57mm]{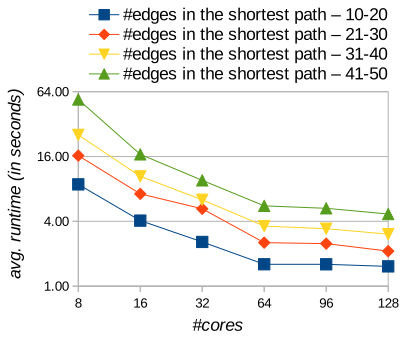}}
\quad
\subfigure[comparison of runtime in London road network.  Y-axis is in $log_4$ scale.]{\label{fig:CPU-utilization-c}\includegraphics[height=45mm,width=57mm]{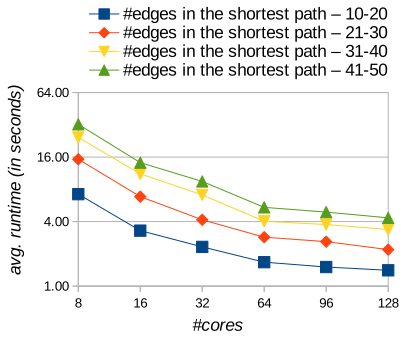}}

\caption{Evaluating Parallel-Spatial-RG for different number of threads. Overhead=$30\%$, $40\%$ edges with non-zero score values and recursion depth $\theta=1$.}
\label{fig:CPU-utilization}
\vspace{-5mm}
\end{figure*}

In this experiment, the execution time of \emph{Parallel-Spatial-RG} is analyzed for different number of cores ($8, 16, 32, 64, 96, 128$) and for different path length ranges ($11$--$20$, $21$--$30$, $31$--$40$, $41$--$50$). As per Fig. \ref{fig:CPU-utilization}, \emph{Parallel-Spatial-RG} gives an almost linear scale up with the increasing number of cores. This shows the scalability of \emph{Parallel-Spatial-RG} algorithm. Though we could not run \emph{Parallel-Spatial-RG} on a bigger system due to resource constraints, the scalability demonstrated by \emph{Parallel-Spatial-RG} guarantees a better performance on a system with a higher number of cores. Hence, we fix the number of cores to 128 for the later part of the experiments.

In the Delhi road network \emph{Parallel-Spatial-RG} had a maximum runtime of $1.32$ seconds and $0.81$ seconds on an average across all the path length ranges. For both the Buenos Aires and London road networks, the performance of \emph{Parallel-Spatial-RG} is quite similar as both have nearly the same network size. For the Buenos Aires dataset, \emph{Parallel-Spatial-RG} had a runtime of $2.7$ seconds on average and $4.2$ seconds for a maximum case. And for the London network, those values were $2.9$ seconds and $4.15$ seconds, respectively. 

To summarize, the proposed method \emph{Parallel-Spatial-RG} produces high solution quality within an acceptable time limit for different datasets. For the bigger networks, it incurs a little bit higher execution time (more than $3$ seconds). This is because of the multiple numbers of shortest path computation on-the-fly for a single CPO problem instance. In our implementation, we used A* algorithm (with Euclidean distance as the heuristic function) for computing shortest paths. 

One may trivially improve the performance of \emph{Parallel-Spatial-RG} by using the latest shortest path computation techniques such as \cite{phast,jing1998hierarchical,sanders2005}. However, experimental evaluation with these techniques is beyond the scope of this paper and will be investigated in the future. An interesting fact to note is that in all the datasets, for the shorter path lengths($11$--$20$), the improvement with the increase in the number of cores gets saturated. This is due to the following two reasons: (a) runtime was already in around one second; (b) the number of sub-tasks created during the run was less than the total number of cores provided for the run. 

\begin{figure}[ht]
\centering
\subfigure[run-time comparison for London road network]{\label{fig:budget-time-c}\includegraphics[width=40mm,height=35mm]{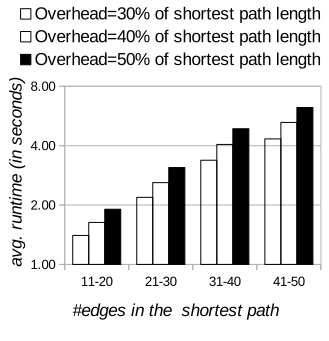}}
\quad
\subfigure[score-gain comparison for London road network]{\label{fig:budget-score-c}\includegraphics[width=40mm,height=35mm]{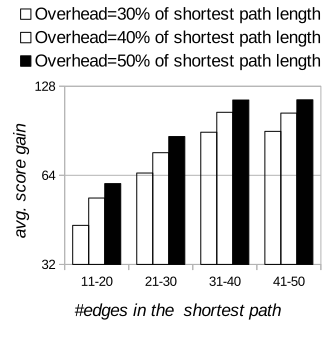}}
\caption{Comparing runtime of Parallel-Spatial-RG for $30\%$, $40\%$, and $50\%$ overhead over the shortest path cost in the road network. $40\%$ of the edges has a non-zero score value and recursion depth $1$. Y-axis is in $log_2$ scale.}

\label{fig:budget}
\end{figure}

 \subsubsection{Varying overhead over the cost of the shortest path:}
\label{sec:budget}
The performance of \emph{Parallel-Spatial-RG} depends on the budget value. To this end, we analyzed the behavior of \emph{Parallel-Spatial-RG} with varying overhead values of $30\%$, $40\%$, and $50\%$. Note that budget = cost of the shortest path + overhead over the shortest path cost. Fig. \ref{fig:budget} reveals that the runtime, as well as the solution quality of \emph{Parallel-Spatial-RG}, increases with the increase in overhead, i.e., increase in budget. As the behavior of \emph{Parallel-Spatial-RG} is quite similar across the datasets, we only include the results for the London dataset for this experiment and fixed the overhead value to $30\%$ over the cost of the shortest path for the later parts of the experiments.

\begin{figure}[ht]
\centering
\subfigure[run-time comparison for London road network]{\label{fig:nav-edges-time-c}\includegraphics[width=40mm,height=35mm]{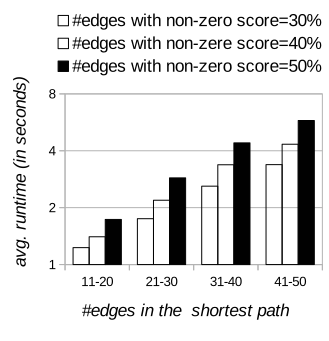}}
\quad
\subfigure[score-gain comparison for London road network]{\label{fig:nav-edges-score-c}\includegraphics[width=40mm,height=35mm]{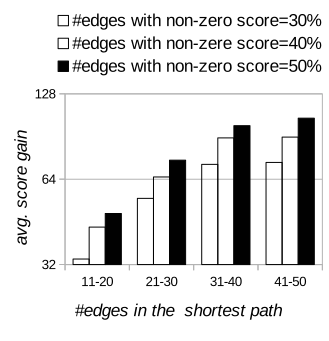}}
\caption{Comparing runtime of Parallel-Spatial-RG for $30\%$, $40\%$, and $50\%$ edges with non-zero scores values in the road network and overhead of $30\%$ over the shortest path cost and recursion depth $1$. Y-axis is in $log_2$ scale.}

\label{fig:nav-edges}
\end{figure}

  \subsubsection{Varying number of edges with non-zero scores values in the road network:}
\label{sec:nav-edges}
The performance of \emph{Parallel-Spatial-RG} also depends on the density of the edges with non-zero scores values. Therefore, we analyzed its behavior with varying density of the edges with non-zero score values ($30\%$, $40\%$, and $50\%$). 
Fig. \ref{fig:nav-edges} demonstrated that the runtime and the solution quality of \emph{Parallel-Spatial-RG} increase with the increase in density of edges with non-zero scores values. Here also, we only consider the results for the London dataset and fixed the density of edges with non-zero score value to $40\%$ for the later parts of the experiments.

\begin{figure*}[ht]
\centering
\subfigure[average score comparison for Delhi road network]{\label{fig:candidate-a}\includegraphics[width=57mm,height=35mm]{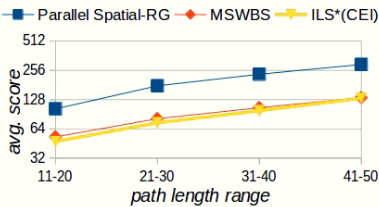}}
\quad
\subfigure[average score comparison for Buenos Aires road network]{\label{fig:candidate-b}\includegraphics[width=57mm,height=35mm]{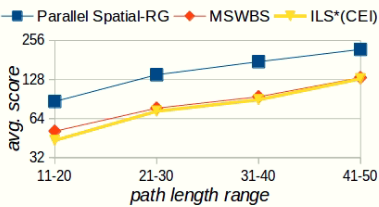}}
\quad
\subfigure[average score comparison for London road network]{\label{fig:candidate-c}\includegraphics[width=57mm,height=35mm]{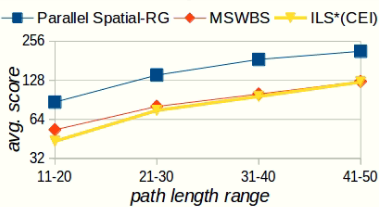}}

\caption{Comparison of Parallel-Spatial-RG,  MSWBS, and ILS*(CEI) for $30\%$ overhead over the shortest path cost, $40\%$ of edges with a non-zero score, and recursion depth $1$. Y-axis is in $log_2$ scale.}
\label{fig:candidate}
\end{figure*}
\subsection{Comparison with Candidate Algorithms}
\label{sec:MSWBS-vs-pspatial}
This experiment compares \emph{Parallel-Spatial-RG} with the candidate algorithms  \emph{MSWBS} and \emph{ILS*(CEI)} algorithm for various path length ranges ($11$--$20$, $21$--$31$, $31$--$40$, $41$--$50$) on each of our three road network datasets. Fig. \ref{fig:candidate} shows the average score comparison between \emph{Parallel-Spatial-RG}, \emph{MSWBS}, and \emph{ILS*(CEI)}. In terms of achieved score, \emph{Parallel-Spatial-RG} outperforms both \emph{MSWBS} and \emph{ILS*(CEI)} by a huge margin in all the three datasets.

With regards to execution time, \emph{MSWBS} had an average (across different path lengths) run-time of $0.12$sec, $0.13$sec, and $0.11$sec on Delhi, Buenos Aires, and London datasets respectively.  Whereas, \emph{Parallel-Spatial-RG} demonstrated an average (across different path lengths) runtime of $0.9$ seconds, $2.7$ seconds, and $2.9$ seconds respectively for the same parameters. \emph{ILS*(CEI)} allows us to fix the execution time for each query and return the obtained path up to that particular threshold. In our experiments, we fixed this threshold to be $3$ seconds (as \emph{Parallel-Spatial-RG} had a maximum execution time of $3$ seconds) in this experiment. Note that \emph{Parallel-Spatial-RG} could have obtained a much lower execution time on a system with more available threads due to ``almost'' linear scale-up. \emph{MSWBS} already has a running time in milliseconds, and \emph{ILS*(CEI)} has an almost fixed runtime (due to its threshold). Therefore, due to lack of space, we didn't include the runtime comparison between \emph{Parallel-Spatial-RG} with the candidate algorithms.

\section{Conclusion}
\label{con}
This paper studied the problem of the Constrained Path Optimization (CPO) problem on road networks. CPO problem has value addition potential in the domain of urban navigation. However, the current state-of-the-art solutions (approximation algorithm or heuristic solutions) either fail to scale up to real-world road networks or have poor solution quality. In contrast, our proposed parallel algorithm \emph{Parallel-Spatial-RG} shows promising results in terms of both scalability and solution quality. In the future, we will continue working on the \emph{Parallel-Spatial-RG} algorithm to improve its scalability even further and also establish a formal approximation ratio for the algorithm. More specifically, we plan to explore the potential of hierarchical routing techniques for improving the scalability of our \emph{Parallel-Spatial-RG} algorithm. 

\bibliographystyle{IEEEtran}
\bibliography{bibliography}

\end{document}